\documentclass[sigconf,nonacm]{acmart} 
\renewcommand
\footnotetextcopyrightpermission[1]{}
\AtBeginDocument{%
  }
\usepackage{todonotes}
\usepackage{hyperref}
\usepackage{amsfonts} 

\usepackage{amssymb}
\usepackage{tabularx}
\usepackage{booktabs}
\usepackage{caption}
\usepackage{minted}

\setcopyright{none}
\usepackage{fvextra}
\DefineVerbatimEnvironment{MyVerbatim}{Verbatim}{
  breaklines=true,
  breakanywhere=true,
  fontsize=\small
}

\usepackage{soul} 
\usepackage{ulem} 
\newif\ifhighlight
\highlightfalse 

\newcommand{\hltext}[1]{%
    \ifhighlight
            \begingroup
            \hl{#1}%
            \endgroup
    \else
        #1
    \fi
}

\renewcommand\footnotetextcopyrightpermission[1]{} 
\begin{document}

\title[Exploring AI-Augmented Sensemaking of Patient-Generated Health Data]
{Exploring AI-Augmented Sensemaking of Patient-Generated Health Data: A Mixed-Method Study with Healthcare Professionals in Cardiac Risk Reduction}

\author{Pavithren V. S. Pakianathan}
\affiliation{%
  \institution{Ludwig Boltzmann Institute for Digital Health and Prevention}
  \city{Salzburg}
  \country{Austria}
}
\affiliation{%
  \institution{LMU Munich}
  \city{Munich}
  \country{Germany}
}

\author{Rania Islambouli}
\affiliation{%
  \institution{Ludwig Boltzmann Institute for Digital Health and Prevention}
  \city{Salzburg}
  \country{Austria}
}

\author{Diogo Branco}
\affiliation{%
  \institution{LASIGE, Faculdade de Ciências, Universidade de Lisboa}
  \city{Lisbon}
  \country{Portugal}
}

\author{Albrecht Schmidt}
\affiliation{%
  \institution{LMU Munich}
  \city{Munich}
  \country{Germany}
}

\author{Tiago Guerreiro}
\affiliation{%
  \institution{LASIGE, Faculdade de Ciências, Universidade de Lisboa}
  \city{Lisbon}
  \country{Portugal}
}

\author{Jan David Smeddinck}
\affiliation{%
  \institution{Ludwig Boltzmann Institute for Digital Health and Prevention}
  \city{Salzburg}
  \country{Austria}
}



\begin{abstract}
Individuals are increasingly generating substantial personal health and lifestyle data, e.g. through wearables and smartphones. While such data could transform preventative care, its integration into clinical practice is hindered by its scale, heterogeneity and the time pressure and data literacy of healthcare professionals (HCPs). We explore how large language models (LLMs) can support sensemaking of patient-generated health data (PGHD) with automated summaries and natural language data exploration. Using cardiovascular disease (CVD) risk reduction as a use case, 16 HCPs reviewed multimodal PGHD in a mixed-methods study with a prototype that integrated common charts, LLM-generated summaries, and a conversational interface. Findings show that AI summaries provided quick overviews that anchored exploration, while conversational interaction supported flexible analysis and bridged data-literacy gaps. However, HCPs raised concerns about transparency, privacy, and overreliance. We contribute empirical insights and sociotechnical design implications for integrating AI-driven summarization and conversation into clinical workflows to support PGHD sensemaking.
\end{abstract}


\begin{teaserfigure}
  \includegraphics[width=\textwidth]{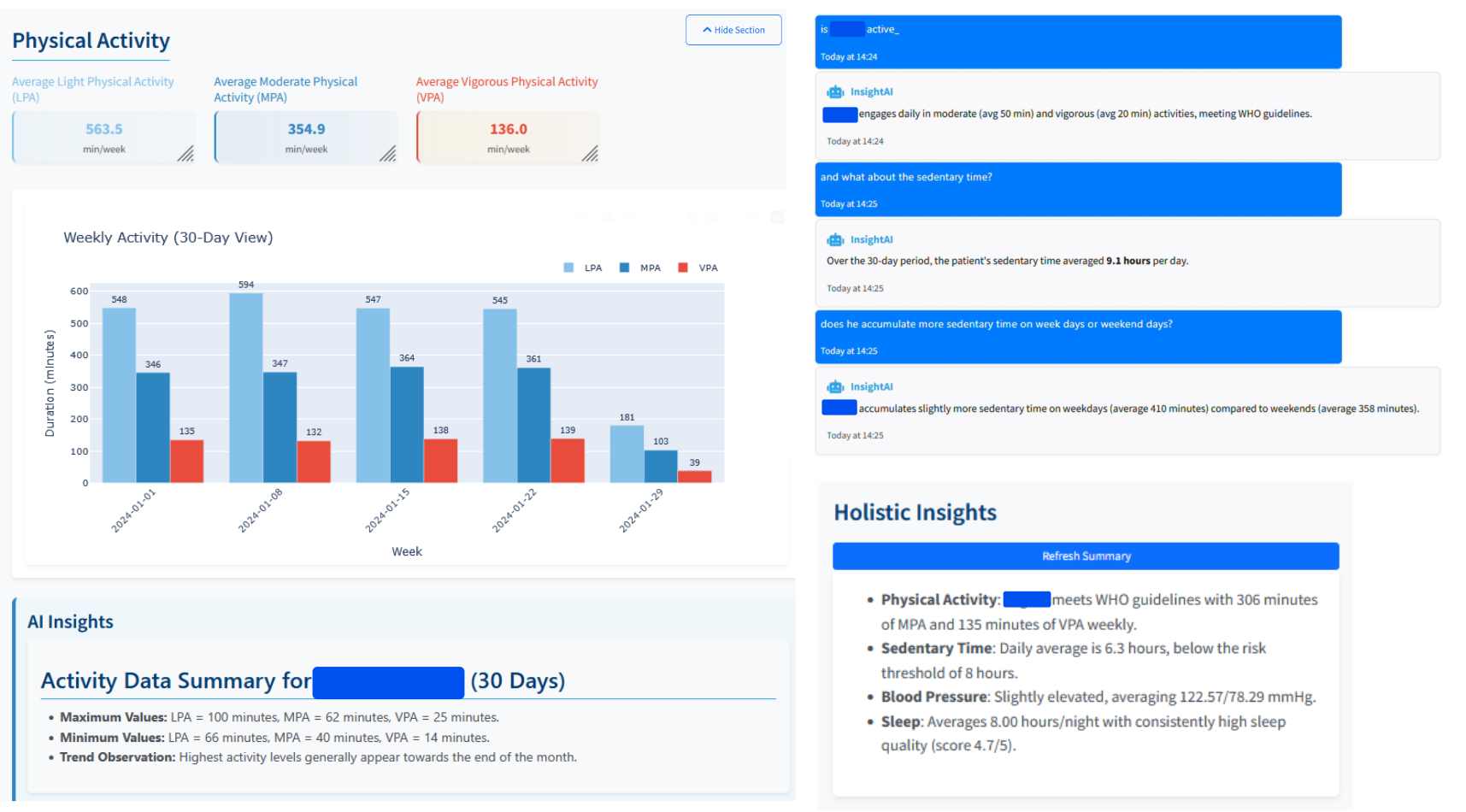}
  \caption{Screenshots from the interface used in the study: (left) Dashboard view showing weekly physical-activity minutes with accompanying LLM-generated summaries. (top right) Example chart from participant P16 using the conversational interface to explore patient-generated health data.  (bottom right) Holistic Insights showing a summary of the persona's physical activity, sedentary time, blood pressure and sleep during a selected time period. (Persona name is blinded for review.)}
  \Description{ Screenshots from the interface used in the study: (left) Dashboard view showing weekly physical-activity minutes with accompanying LLM-generated summaries. (top right) Example chart from participant P16 using the conversational interface to explore patient-generated health data.  (bottom right) Holistic Insights showing a summary of the persona's physical activity, sedentary time, blood pressure and sleep during a selected time period. (Persona name is blinded for review.)}
  \label{fig:teaser}
\end{teaserfigure}



\maketitle

\section{Introduction}
\label{sec:introduction}
The widespread adoption of wearables and digital sensing tools like smartwatches, smart weighing scales, and blood pressure monitors have led to the proliferation of patient-generated health data (PGHD) \hltext{--} health and lifestyle-related data that individuals collect outside traditional clinical settings, typically through wearables, smartphone apps, or home-use medical devices (e.g. physical activity minutes, sedentary time, heart rate, sleep, blood pressure) which can support preventative care \cite{tadas_using_2023}. PGHD is becoming more multimodal \cite{friedewald_ubiquitous_2011} and increasingly extends beyond traditional clinical or health practice settings, playing a key role in enabling pervasive healthcare. PGHD offers healthcare professionals (HCPs) a more holistic and objective view of their patients, enabling personalized care. 

However, integrating PGHD into clinical workflows remains challenging. At the data level, PGHD is heterogeneous, often stored in non-standardized formats and lacks interoperability with clinical systems \cite{tadas_using_2023,west_common_2018,pakianathan_barriers_2024}. At the clinician level, the sheer volume and heterogeneity of multimodal PGHD, often accumulated over months, impose significant sensemaking demands - i.e., the effortful process of interpreting, synthesizing, and drawing actionable insights from complex and heterogeneous data - on already overburdened HCPs \cite{dunn_what_2025}. The growing amount of data work, collecting, interpreting, and integrating multimodal PGHD and EHR information, places heavy cognitive demands on HCPs, intensifies time pressure, and can lead to burnout \cite{adler_burnout_2022,budd_burnout_2023,ye_impact_2021}. In addition, many HCPs face challenges with visualization literacy, the ability to correctly interpret and critically evaluate graphical representations of data~\cite{nobre_reading_2024,dunn_what_2025}. Such gaps risk undermining clinical decision-making~\cite{weil_visualization_2022}.

To address these barriers, researchers are exploring AI - particularly large language models (LLMs) - to generate summaries and natural language explanations that augment sensemaking. The integration of AI to support and augment HCPs in their workflows has been broadly investigated for several years \cite{maslej_out_2023,zajac_clinician-facing_2023}. The rapid advancement of LLMs has led to increasing research interest in their application to clinical data interpretation. In particular, recent studies have examined their use in supporting health data sensemaking through automated summaries and conversational interface narratives \cite{li_vital_2025,stromel_narrating_2024} in non-clinical contexts. 

Despite \hltext{the recent commercial release of LLM-based tools to support healthcare organizations} \cite{openai_introducing_2026,google_healthAI_2026,aws_healthAI_2026} and patients \cite{chatgpt_health_2026}, effective real-world implementation and adoption of clinical AI remains challenging, especially in the \textit{last mile} \cite{coiera_last_2019}. There remain several open questions regarding how LLM-based systems can be better aligned with HCP workflows for better integration within a broader sociotechnical context. 
Recent work on ML-based healthcare systems has prompted a re-framing of Human-AI interaction as the design of sociotechnical ML systems \cite{zajac_clinician-facing_2023}.

We use cardiovascular risk reduction as a use case to explore LLMs for augmenting data sensemaking. Cardiovascular disease \hltext{(CVD)} is the largest cause of death worldwide \cite{who_cardiovascular_nodate}. Healthier behaviours (e.g. increased physical activity, a prevention measure for cardiac conditions) and staying within safe zones (e.g. controlled blood pressure) can reduce risk of cardiac disease \cite{bull_world_2020,mcgowan_exploring_2024,pinckard_effects_2019}.

In cardiac risk reduction, patients or at-risk individuals may self-track vital signs such as ECGs, blood pressure, sleep, and physical activity patterns, capturing rich multimodal data representations of themselves. \hltext{During physical activity planning sessions, such PGHD, which are often already collected as part of patients' daily routines, can provide HCPs with a more holistic and objective view of their patients, effectively serving as a digital twin} \cite{katsoulakis_digital_2024,thangaraj_cardiovascular_2024} that can guide and inform care. This information, in turn, enhances risk assessment and medical decision making \cite{pedroso_leveraging_2025}.

We position LLM summaries and conversational interfaces as potential \textit{cognitive sensemaking partners}, extending the theory of distributed cognition by Hutchins \cite{hutchins_cognition_1995} and work on sensemaking in HCI \cite{russell_cost_1993,pirolli_sensemaking_2005} by exploring in our use case how AI outputs can externalize interpretive effort and reshape HCPs' data reasoning strategies. Furthermore, we explored HCP perspectives on workflow integration, perceived benefits, limitations, and acceptability. 

\label{sec:rq}
We conducted a mixed-methods study with 16 HCPs with two parts. 
Our study addresses the following research questions (RQs):
\begin{enumerate}
    \item How do LLM-generated summaries affect HCPs’ ability to make sense of multimodal PGHD?
    \item What are the perceived benefits, limitations, and acceptability of integrating LLM summaries into clinical workflows?
    \item How do HCPs perceive and use conversational interfaces to explore and analyze PGHD and augment data-enabled workflows?
\end{enumerate}

In the first part, \hltext{HCPs} reviewed one year of synthetic multimodal data representing individuals at risk for cardiovascular disease, both with and without LLM-generated summaries. In the AI condition, each chart was accompanied by a summary as well as an overall multi-modal summary. In the second part, they used a conversational interface for data analysis and sensemaking. We concluded with semi-structured interviews to capture their perspectives on LLM integration. \hltext{In line with established approaches at early technology-readiness levels, we position this work as formative and exploratory. Our aim is not to evaluate clinical effectiveness, but to generate insights into acceptable and effective interaction patterns around AI-generated summaries and conversational interfaces for supporting sensemaking. By researching how HCPs perceive such tools and what sociotechnical considerations shape their potential integration into practice, the study is intended to surface opportunities, risks, and design implications rather than to make claims about clinical performance or real-world outcomes.}

Our findings show that HCPs valued LLM-generated overviews for reducing the time required to make sense of complex data. They also recognized the potential of conversational interfaces to augment sensemaking and address gaps in data literacy. At the same time, they emphasized the need for more personalized summaries tailored to the data subject and expressed concerns about transparency, privacy, overreliance, and possible deskilling.
\label{sec:contribution}
Our study contributes to a growing body of literature in HCI and Human-AI interaction research by surfacing the potential sociotechnical implications of introducing LLMs into PGHD-driven clinical practice. Specifically, we:
\begin{enumerate}
    \item Provide empirical insights from a mixed-methods evaluation of HCPs’ perceptions, usability, workload, trust, and acceptance of LLM-generated summaries for multimodal PGHD.
    \item Investigate conversational interfaces for PGHD exploration, extending dashboards with natural language interaction to support data literacy and augment workflows.
    \item Provide a sociotechnical understanding of LLM integration in clinical practice, including design implications around provenance-aware summaries, personalization, transparency, data governance, and strategies to mitigate deskilling.
\end{enumerate}

\section{Related Work}
\label{sec:RW}
Research on PGHD has examined its benefits and challenges, explored how HCPs interpret complex personal health data and, more recently, considered how AI can support interpretation through automated summaries and conversational interfaces. 
However, these bodies of work remain largely disconnected: studies on PGHD integration rarely consider AI support, while evaluations of LLMs rarely examine real-world clinical workflows. 
Our work bridges these areas by investigating how LLM-generated summaries and natural language interfaces can support HCPs in making sense of multimodal PGHD and augment data-enabled clinical workflows.
\subsection{Challenges of Multimodal, Longitudinal PGHD in Clinical Care}
Advances in sensor miniaturization and cost reduction, along with widespread adoption of consumer-ready wearables and home-use devices, have enabled the collection of large volumes of PGHD. Such PGHD has the potential to bridge information gaps about patients~\cite{guardado_use_2024}. PGHD has shown promise for managing lifestyle-related conditions such as diabetes, obesity, and cardiovascular disease. Yet integrating PGHD into care pathways remains challenging due to interoperability issues, privacy and trust concerns, and the lack of standardized data formats or effective sensemaking tools\hltext{ }based on heterogeneous data streams, representation formats, sampling frequencies, data ranges, etc., leading to gaps in implementation~\cite{pakianathan_barriers_2024,west_common_2018}

For example, in the context of irritable bowel syndrome, patients who self-track their food and symptom data face challenges when sharing them with HCPs, such as insufficient time to review large amounts of data and having personalized and actionable recommendations \cite{chung_identifying_2019}. Researchers try to address such limitations by building infrastructures that support data sharing with HCPs. In a study focused on Type 2 Diabetes, researchers developed the DiaFocus system \cite{bardram_diafocus_2023} to investigate the collection of physiological, behavioral, and contextual data in combination with ecological assessments of psycho-social factors to facilitate management of Type 2 diabetes in primary care and serve as a dialogue tool between patients and HCPs.

In cardiovascular care, PGHD such as ECG, physical activity, blood pressure, and sleep can be highly valuable for assessing risk,  personalizing treatment \cite{nick_e_j_west_personalized_2022,varma_promises_2024,bull_world_2020,tadas_using_2023} \hltext{and heart health monitoring} \cite{keys2024}. \hltext{Recent CHI work have also highlighted the increase in self-tracking to manage CVD} \cite{keys2024} and \hltext{increasing acceptance of PGHD in cardiac care} \cite{keys2025,tadas_using_2023}. However, the volume, heterogeneity, and lack of standardization of multimodal and longitudinal \hltext{PGHD}, often spanning months and including high-frequency streams or co-morbidities, can overwhelm HCPs \hltext{in resource-constrained settings}, increasing cognitive burden and sometimes leading to decision paralysis rather than support \cite{pedroso_leveraging_2025}.

In practice, CVD patients or at-risk individuals may self-track daily blood pressure readings, ECGs, and physical activity patterns (e.g., duration or time spent in a heart rate zone). While these metrics are crucial for HCP decision-making, the “cognitive burden of understanding and synthesizing this information must be minimised and the opportunity to make good decisions maximised” \cite{backonja_data_2018}, especially in time-limited consultations. Simply collecting PGHD does not translate into actionable insights; HCPs require well-designed tools that make longitudinal and multimodal streams useful for decision-making.

\subsection{Sensemaking of Patient Generated Health Data}
Sensemaking is broadly understood as an iterative process of gathering and interpreting information to enable action \cite{klein_dataframe_2007,pirolli_sensemaking_2005,russell_cost_1993}. In healthcare, making sense of PGHD enables HCPs to develop a more holistic and objective view of patients, track recovery, and personalize decision-making. However, in time-limited consultations, the costs of interpreting and exploring longitudinal, multimodal data can be high. As van Wijk notes, visualizations impose perception and exploration costs that may outweigh their benefits when data is complex or unfamiliar \cite{van_wijk_views_2006}. For HCPs, such costs can translate into cognitive overload, “analysis paralysis” \cite{jones_sensemaking_2016}, and increased workload. 

Visualization literacy further complicates this challenge. While HCPs are trained to interpret standardized lab results and clinical charts, consumer-facing PGHD often relies on heterogeneous metrics, visual conventions, and time scales that are less familiar \cite{west_common_2018,dunn_what_2025}. Given the relationship between visualization literacy and decision-making capacity \cite{weil_visualization_2022}, these gaps can hinder effective clinical use of PGHD. 

Recent research suggests AI could alleviate these burdens \hltext{by automating interpretation and synthesis}. Stromel \hltext{\textit{et al.}} \cite{stromel_narrating_2024} used LLMs to generate narratives from fitness tracker data, \hltext{Shaan \textit{et al.}} \cite{shaan2025} \hltext{uncovered LLM's potential in supporting planning and action and interpreting data-driven insights using self-tracked data}, while Li \textit{et al.} \cite{li_vital_2025} showed how LLMs could support expert analysis of multimodal personal tracking data. More broadly, advanc\hltext{ements} in AI have increased its ability to analyze and generate insights from complex datasets \cite{tu_what_2024} \hltext{and recent HCI work in the domain of mental health have also shown promises of LLMs in generating natural language summaries synthesizing multi-modal data to create clinical insights} \cite{Englhardt24,kim2024} \hltext{and enable human-AI collaborative decision-making} \cite{van2025}. \hltext{These advancements} create potential to bridge literacy gaps and offload cognitive effort in sensemaking. Yet most work evaluates LLMs in isolation; little is known about how AI-generated summaries might complement visualizations, reduce cognitive burden, or fit within \hltext{cardiac risk reduction} clinical workflows \cite{backonja_data_2018,feller_visual_2018,scholich_augmenting_2024}. A nuanced understanding of how AI can augment PGHD sensemaking \hltext{in such workflows from the perspectives of HCPs} is therefore essential.

\subsection{AI Augmented (Health) Data Sensemaking}
AI is increasingly being implemented in clinical settings to support decision-making and augment HCP workflows \cite{zajac_towards_2025}. Recent advances in LLMs have opened new opportunities for sensemaking of complex, multimodal data. Researchers have explored their use to \hltext{summarize unstructured data from electronic health records} \cite{ahsan2024retrieving,fraile2025understanding} \hltext{- with emerging evidence of LLMs outperforming medical experts in text summarization} \cite{van2024adapted}, narrate personal tracking data \cite{stromel_narrating_2024}, support ubiquitous computing experts in analyzing multimodal data \cite{li_vital_2025,vs_pakianathan_towards_2024}, and even finetune models (e.g., SensorLM) to translate wearable data into human-readable narratives \cite{zhang_sensorlm_2025}. In the context of distributed cognition, such tools and systems act as \textit{cognitive sensemaking partners}, helping individuals offload effort when interpreting data \cite{hollan_distributed_2000, grinschgl_supporting_2022}.

In cardiovascular contexts, AI shows particular promise for prevention and monitoring. Reviews highlight its role in summarizing PGHD to reduce overload and enable more personalized care \cite{aminorroaya_harnessing_2025,croon_emerging_2025,pedroso_leveraging_2025}. The American Heart Association has also called for digital cardiac rehabilitation tools to prioritize automated interpretation using AI \cite{golbus_digital_2023}. Early explorations of generative AI in health and sports science have also examined applications such as exercise planning, where LLMs can draft training programs but still require experts in the loop \cite{dergaa_using_2024,farquhar-snow_artificial_2025}.  

Despite this promise, challenges remain. HCPs raise concerns about overreliance, accuracy, data privacy, trust, workflow integration, and loss of autonomy \cite{henzler_healthcare_2025,lambert_integrative_2023,sahoo_health_2025} when integrating AI. Oversimplified natural language outputs can mislead or be misaligned with user goals \cite{scholich_augmenting_2024,jones_sensemaking_2016}. Acceptance also varies generationally: younger, more digitally fluent HCPs often view AI tools as helpful extensions of their practice, while older or less tech-comfortable colleagues may resist integration with misalignment with established practices \cite{a_s_adoption_2024}. Broadly, successful implementation requires situating AI within sociotechnical contexts, ensuring that AI systems align with existing work practices, infrastructures, and social relations \cite{zajac_clinician-facing_2023}. 

\subsection{Research Gap and Contribution}

Although PGHD dashboards and LLM-generated summaries have been studied independently, little is known about how their integration can support clinicians’ sensemaking in real-world, time-pressured workflows. Existing evaluations rarely situate LLM tools within the sociotechnical contexts of clinical practice, leaving unanswered questions about workflow fit, trust, and responsible adoption. Addressing this gap is essential to unlock PGHD’s potential for preventative and health care. This sociotechnical alignment gap is where our work contributes.
Building on research that frames analytic technologies as external scaffolds in distributed cognitive systems \cite{hollan_distributed_2000,norman_cognitive_1991}, our study extends this perspective by examining how AI-generated summaries and conversational interfaces function as \textit{cognitive sensemaking partners} in the clinical interpretation of multimodal health data.

\section{Method}
\label{sec:method}
\begin{figure}
    \centering
    \includegraphics[width=1\linewidth]{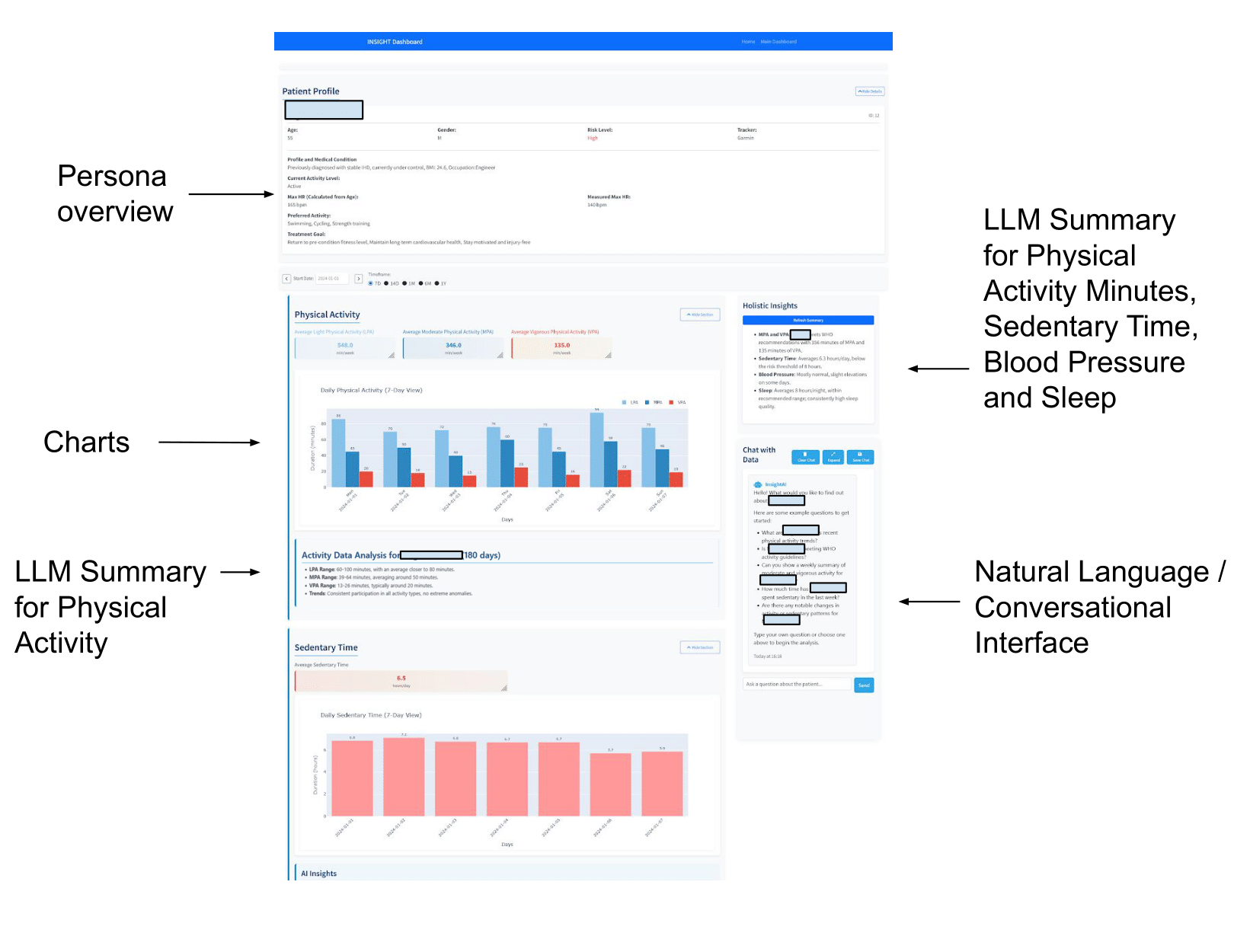}
    \caption{A snippet of the interface in demo mode showing an overview of the persona, charts of physical activity minutes and corresponding LLM summaries, holistic insights of physical activity minutes, sedentary time, blood pressure and sleep, and charts of sedentary time over a one-week period. Summaries for sedentary time, and other information about blood pressure and sleep and draft physical activity plan are not visible. (Persona name is blinded for review)}
    \label{fig:interface}
\end{figure}

We employed a mixed-methods within-subjects research design to investigate how LLM-generated summaries influence HCPs’ sensemaking of multimodal PGHD in the context of cardiac risk reduction and physical activity planning. We also explored HCP perspectives on LLM-enabled conversational interface data exploration and analysis through a prototype concept embedded within our data summary visualization and exploration interface (Figure \ref{fig:interface}).

\subsection{Study Apparatus}
\label{sec:studyApparatus}
We developed a custom dashboard to simulate multimodal PGHD review in cardiovascular care and physical activity planning (Figure \ref{fig:interface}). The dashboard presented synthetic multimodal PGHD, generated as described in \hyperref[sec:synthData]{Synthetic Data Generation} section. The dashboard design was informed by a pre-study with six self-trackers and five HCPs (anonymized paper submitted as supplementary material). \hltext{Furthermore, we chose a one-year time horizon for the synthetic data based on consultation with two HCPs in the pre-study, who confirmed that longitudinal overviews spanning several months to a year are clinically meaningful for cardiovascular risk reduction, despite variation in actual follow-up intervals across services.} These iterative steps informed key interface features and data visualization requirements. The system was implemented in  Python using the Plotly dash framework\footnote{https://dash.plotly.com/}, a low-code framework for building data apps in Python. Data about patient personas, physical activity plans and the synthetic PGHD were stored in Comma-Separated Values (CSV) files.

\begin{figure}
    \centering
    \includegraphics[width=0.7\linewidth]{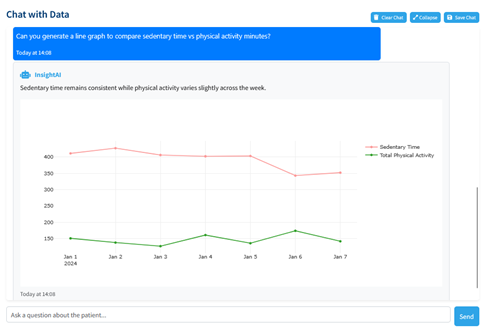}
    \caption{Expanded view of the conversational interface with response for a query asking to compare sedentary time with physical activity minutes using a graph.}
    \label{fig:chatbot}
\end{figure}

The system supported two conditions: AI Summary and No-AI Summary. 
In the \textbf{AI Summary condition}, LLM-generated bullet-point summaries integrated persona characteristics with data from four modalities (physical activity, blood pressure, sleep, sedentary time). 
Summaries highlighted key trends, anomalies, and comparisons to WHO recommendations \cite{who_cardiovascular_nodate}, and were refreshed based on user-selected time frames (e.g., 1 week, 2 weeks, 1 month, 6 months, 1 year). \hltext{We selected WHO physical activity recommendations as benchmarks because HCPs in our context routinely reference them and they align with national guidance, ensuring clinical familiarity.} Prompts for generating the summaries were iteratively refined with HCP input to ensure clinical relevance (see \hyperref[app:prompt]{Appendix} for details). In the \textbf{No-AI Summary condition}, participants viewed the same charts without AI summaries. After completing both conditions, participants interacted with a conversational interface (Figure \ref{fig:chatbot}) for \hltext{natural-language data exploration and on-demand chart generation. HCPs could use natural-language to compare variables (e.g., sedentary time vs. physical activity), describe trends, or highlight correlations across modalities. For visualization requests, the LLM was prompted to return a Plotly JSON specification corresponding to a suitable chart type. The server parsed this JSON and rendered the resulting visualization directly in the chat interface. During this phase, the system only supported exploratory analysis and visualization queries. It did not provide clinical recommendations, generate exercise prescriptions, or suggest medical actions; such outputs were intentionally disabled. The full conversational prompting schema is provided in the supplementary material.}

\hltext{To assess whether LLM-generated statistics in the model responses would likely be assessed to be believable, we performed a post-hoc provenance analysis of the summaries and the chat transcripts after the study. Using the same model, parameters, and synthetic data for all seven personas, we generated 35 holistic insight summaries across the five duration options in the interface. We then compared LLM-derived averages for activity, sleep, and blood pressure metrics against ground-truth values. The summaries closely approximated actual data with some variability in physical activity minutes but strong consistency in sedentary time and blood pressure. The overall mean absolute percentage difference (MAPD) of 3.96\% (median 0.47\%) indicates that they are broadly realistic. We additionally checked the correctness of ranges and actual values presented in the summaries, and all 25 ranges appearing across the 10 sampled summaries were accurate. These covered physical activity minutes, blood pressure, and sleep duration, suggesting reliable range generation despite incomplete coverage across metrics. Next, we analysed the conversational chat logs from the 16 participants. Across 33 instances where the LLM presented averages (e.g., sedentary time, activity minutes, blood pressure, sleep duration/quality), the MAPD was 2.68\% (median 1.57\%), and no discrepancies were observed for trends or minimum/maximum values. The full modality-specific breakdown (184 averages and 30 averages across physical activity, sedentary time, blood pressure, and sleep metrics for holistic insights and chat logs respectively) is provided in Appendix.}

\subsection{Synthetic Data Generation}
\label{sec:synthData}
We synthetically generated seven personas (one for interface familiarization in demonstration mode and six for the experimental tasks) together with corresponding PGHD. Persona design considered cultural and contextual relevance, drawing on prior work that used personas to support just-in-time adaptive interventions for physical activity \cite{haag_last_2025}. Cardiovascular risk levels were stratified using the SCORE2 criteria \cite{score2_2021}, resulting in three levels: moderate, high, and very high risk. 
Each persona’s metadata file included comorbidities, activity levels, preferred physical activities, medical conditions, BMI, occupation, and physical characteristics. 

From this metadata, and building on an existing dataset by Henriksen \textit{et al.} \cite{henriksen_replication_2023}, we generated one year of multimodal PGHD across four modalities: physical activity, blood pressure, sleep, and sedentary time. \hltext{The duration of the data was informed by recommendations from HCPs in the pre-study.}
Data were produced with ChatGPT (GPT-4-Turbo) by supplying persona characteristics and generating Python-based randomization functions; technical details of prompts and parameters are provided in the Appendix. 
The choice of modalities was informed by HCP input in a pre-study (anonymized paper submitted as supplementary material) as clinically relevant to cardiovascular risk reduction.

In addition, a generic draft physical activity plan was generated with ChatGPT, as prior work has shown such drafts to be sufficient starting points for exercise program design \cite{washif_artificial_2024}. 
These plans were subsequently refined during a pilot session with HCP input to ensure plausibility. 

Use of synthetic PGHD as a methodological choice enables controlled, reproducible study, and ethically low-risk designs for early-stage evaluation, which is particularly appropriate for formative work studying health-data interactions where privacy and security concerns strongly limit data sharing. At the time of research, suitable longitudinal, multimodal PGHD datasets required for the study setup were not available for use, making synthetic data a pragmatic choice for early-stage evaluation. The generated data were verified by the lead researcher to ensure quality and applicability by two experienced HCPs before the study.  While this setup enabled systematic exploration of sensemaking workflows, it also means that findings should be interpreted as reflective of interactions under controlled conditions rather than as evidence of deployment with real-world PGHD. See \hyperref[limit]{Limitations} section for further details.

\subsection{Procedure}
\begin{figure}
    \centering
    \includegraphics[width=0.5\linewidth]{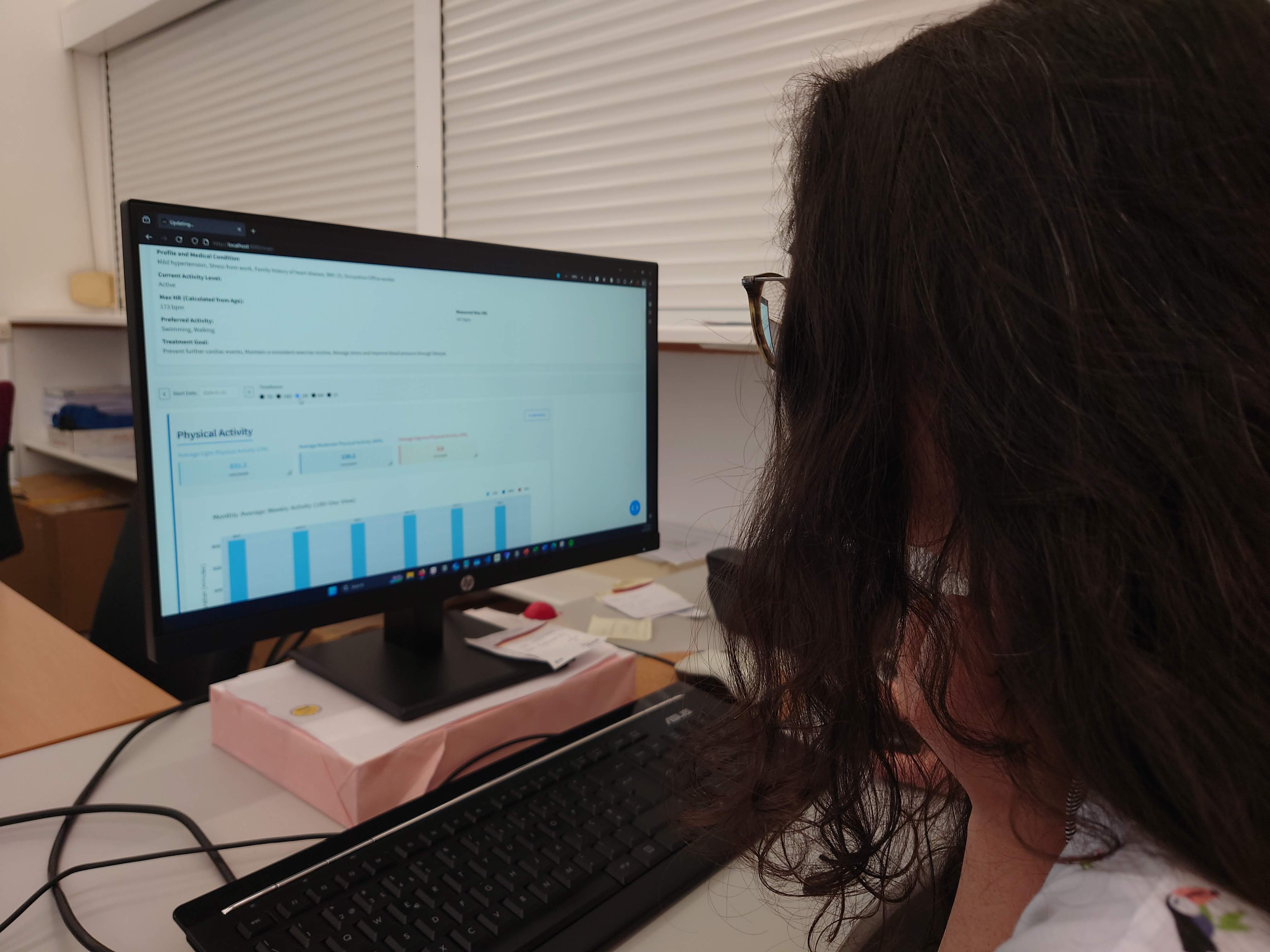}
    \caption{Participant interacting with the interface at their workplace.}
    \label{fig:participantInteraction}
\end{figure}

\begin{figure}
    \centering
    \includegraphics[width=0.7\linewidth]{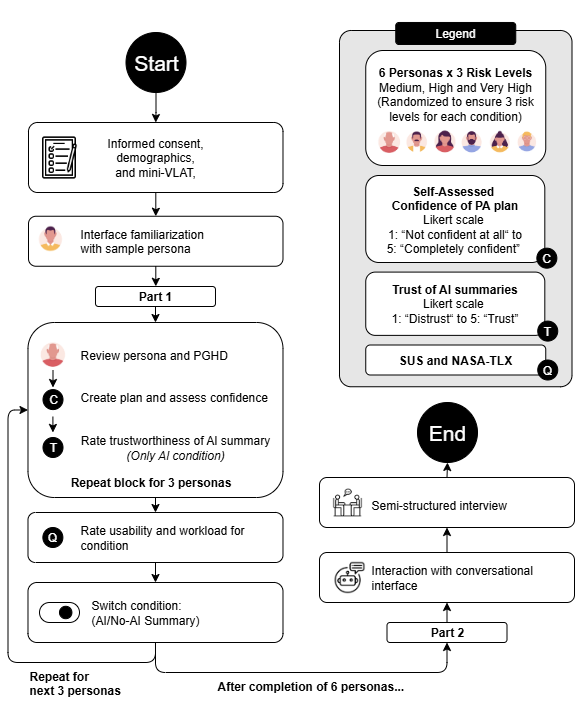}
    \caption{Each session with a HCP was divided into 4 key sections. 1) Informed consent, Demographic Questionnaire, Interface Familiarization 2) Experimental evaluation with AI summary and No AI summary conditions with three randomized persona risk levels and 6 personas in total. 3) Interaction with conversational interface to perform data exploration with natural language. 4) Closing semi-structured interview}
    \label{fig:studyProtocol}
\end{figure}

Each session (c.f Figure \ref{fig:studyProtocol} for detailed procedure) lasted approximately 75 minutes and was conducted in English. \hltext{The study took place in a country where English is not the primary clinical language, but all participants reported being comfortable using English-language research and clinical tools and routinely work with English-language scientific literature and guidelines.} As we were conducting the study with healthcare professionals based in different locations, we conducted the study in their clinical settings (c.f Figure \ref{fig:participantInteraction}\hltext{)} rather than bringing them to a central lab. This approach prioritized ecological validity and \hltext{contextual} realism \hltext{by situating the study in participants’ usual clinical environments} over the consistency of a controlled laboratory environment. \hltext{The study tasks were designed to approximate -- rather than precisely replicate -- time-pressured clinical sensemaking as described by HCPs in prior co-design work. This trade-off between ecological validity and experimental control aligns with our formative research aims} of understanding how LLM summaries and conversational interfaces fit into time-pressured, heterogeneous clinical settings. Thus, findings should be interpreted less as evidence of precise performance differences and more as insights into real-world acceptability and workflow integration.

Participants began with a 15-minute familiarisation phase, including a demographics and MiniVLAT \cite{pandey_mini-vlat_2023} questionnaire. They were then introduced to the system in demo mode using a fixed patient persona. The demo mode allowed participants to fully familiarise with the interface functionality and clarify questions about the interface. It was also to ensure consistent exposure and reduce novelty effects.

The core data-sensemaking task (Part 1, $\approx$ 40 minutes) involved navigating the two  conditions in randomized order. In each block, participants reviewed three unique personas (stratified by CVD risk: medium, high, very high) and were asked to make changes to a pre-defined physical activity plan. \hltext{Participants took $\approx$ 5-7 minutes to review each persona.} After each personas, they rated their confidence in the physical activity plan they created as a result of the decision making process (Likert scale from 1: “Not confident at all“ to 5: “Completely confident”). Additionally in the AI condition, they rated their perceived trust in the AI summary (Likert scale from 1: “Distrust“ to 5: “Trust”). Trust score was framed to participants as their perceived trust in the accuracy, reliability, and clinical appropriateness of the summary. After each condition block (AI/No-AI Summary), they completed NASA TLX \cite{hart_development_1988} and SUS \cite{brooke_sus-quick_1996} questionnaires. 
In Part 2 ($\approx$ 15 minutes), participants interacted with a conversational interface based tool to analyze data for a fixed patient. One participant preferred interacting with the conversational interface in their native language rather than in English. Finally participants completed a 15-minute semi-structured interview. The interview script can be found in the \hyperref[app:interview]{Appendix}. 

\subsection{Participants}
\textbf{HCPs’ Specialties and Expertise.} Sixteen HCPs (12 women, 4 men; \textit{M} = 31.4 years, \textit{SD} = 5.0) specialized in cardiovascular rehabilitation and physical activity counselling (\textit{M}= 9.1 years \textit{SD} = 5.5, range = 2–20 years). Most (n=12) regularly used PGHD in practice. Generative AI use was common for research tasks (e.g., writing, data analysis), with a minority experimenting with clinical support (e.g., exercise prescriptions). Visualization literacy was high (MiniVLAT \textit{M} = 9.47/12). \hltext{None of the HCPs who took part in this study had previously participated in our earlier pre-studies or co-design activities related to this project; this helped us avoid familiarity effects with the prototype and study tasks. We recruited participants via an email publicity sent to HCPs working in cardiac care at a university hospital, which was subsequently shared within their professional networks.}

\textbf{Background with \hltext{PGHD} and tools.} Most participants (n = 12) regularly worked with \hltext{PGHD} for clinical or research purposes, two used PGHD for personal self-tracking, and three reported little or no prior exposure. Fourteen had hands-on experience with commercial data-visualization or analytics platforms (e.g., Excel, SPSS, Power BI, clinical dashboards); one had used a patient-facing fitness application (Polar), and four reported no experience with visualization tools.

\textbf{AI literacy and Adoption.} Generative-AI usage was common amongst HCPs but varied by task: scholarly writing and publication (n = 7, daily); data analysis (n = 6); health-information seeking (n = 5); and exercise-prescription support (n = 2). Three participants indicated that they did “not currently use” AI tools. Thus, while most use cases were academic or research-oriented, a small number of participants reported experimenting with AI for clinical tasks such as tailoring exercise prescriptions. This indicates that HCPs’ AI experience emerged primarily from research exposure, with considerable instances of clinical application.

\textbf{Visualization literacy.} Scores on the MiniVLAT \cite{pandey_mini-vlat_2023} were high (\textit{M} = 9.47, \textit{SD} = 1.09, out of total score of 12) suggesting that HCPs were proficient in interpreting standard visual encodings.

\subsection{Ethics}
Our study protocol received official approval from the relevant institutional ethics committee (blinded for review) prior to data collection. Informed consent was obtained prior to the commencement of the study and permission was obtained for audio recording. Participants were reimbursed 50 EUROs worth of vouchers for their participation.

\subsection{Measures and Analysis Techniques}
We collected both quantitative and qualitative data. Quantitative measures included demographics, MiniVLAT scores \cite{pandey_mini-vlat_2023}, self-assessed confidence (per persona), trust in AI summaries (AI condition only), NASA-TLX workload \cite{hart_development_1988}, and SUS usability \cite{brooke_sus-quick_1996}. Given the small sample and ordinal data, we used Wilcoxon signed-rank tests \cite{wilcoxon_individual_1992} for paired comparisons, Spearman correlations \cite{zar_spearman_2005} for trust–confidence associations, and linear mixed-effects models to account for repeated measures and verify robustness. 

Qualitative data comprised screen recordings, conversational transcripts, and semi-structured interviews on experiences with AI summaries, natural language interaction, and acceptability in workflows. Audio recordings were locally transcribed with OpenAI Whisper \footnote{https://github.com/openai/whisper} and verified by the first author. We analyzed the data using Mayring’s qualitative content analysis \cite{mayring_qualitative_2004}. Two authors co-coded data from four participants to establish consistency, then the first author coded the remaining transcripts, and themes were derived collaboratively. Finally, the codes were grouped into themes reflecting experiences of sensemaking, perceived value, trust, concerns (e.g., privacy, overreliance), and implementation considerations in cardiac care workflows.

\section{Findings}
We present our findings in terms of the value and risks of integrating LLM-generated summaries and conversational interfaces into clinical practice. \hltext{As a formative study, these findings should be interpreted as insights into perceptions, practices, and sensemaking dynamics rather than as evaluations of clinical performance or system correctness.} The first two sections describe how participants saw these tools supporting sensemaking and fitting into their workflows, while the third section highlights perceived risks and tensions that would need to be addressed for responsible adoption. These include questions of autonomy (assistant vs. replacement), trust, overreliance, privacy, and generational acceptance. Together, the findings illustrate the promise of AI to reduce workload and bridge data-literacy gaps, while also surfacing concerns about accountability, professional identity, and sociotechnical alignment.

\subsection{Value of AI Summaries for Sensemaking: Benefits, Evolving Practices, and Emerging Needs}

\subsubsection{One year? It's a lot of data!: Time and effort reduction for data sensemaking}

From our quantitative analysis, \textbf{system usability was very high (A+ range) for both conditions}
\cite{brooke_sus-quick_1996} \textbf{with SUS scores in the AI condition being slightly higher} (mean = 90.63, SD = 8.44) than in the No-AI condition (mean = 85.94, SD = 14.60). We found \textbf{no statistically significant difference in SUS scores between the AI and No-AI conditions}.

\begin{figure}
    \centering
    \includegraphics[width=1\linewidth]{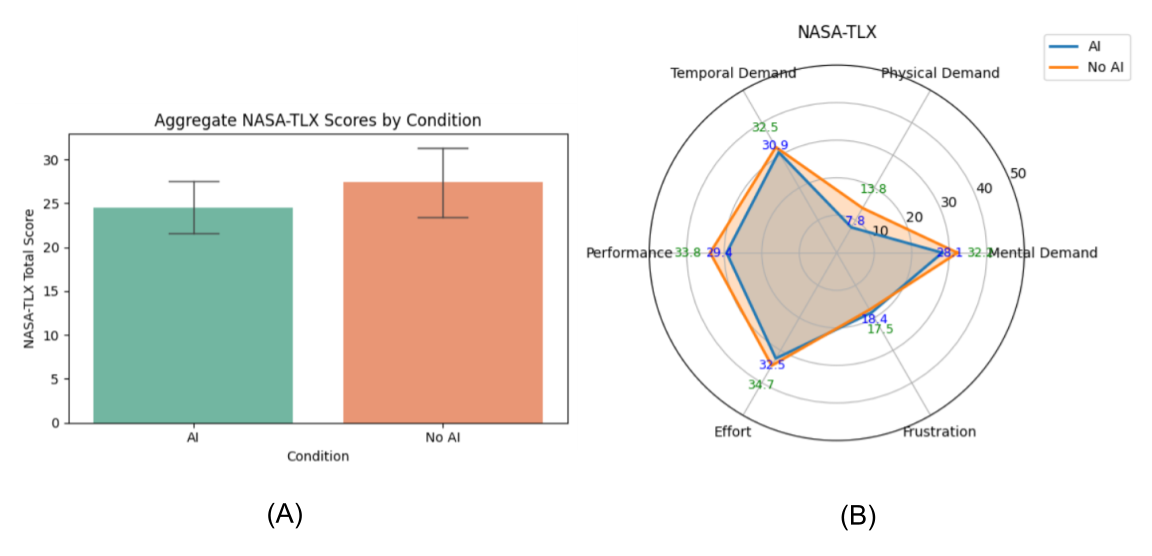}
    \caption{Charts showing the pairwise comparison of NASA-TLX workload ratings for AI and NO-AI summary conditions (n=16).\\
(A) Barchart of Aggregate NASA-TLX Comparison\\
(B) Spider Chart of Subscale NASA-TLX Comparison\\
}
\label{fig:nasaTLX}
\end{figure}

Furthermore, the \textbf{AI summary condition yielded a slightly lower NASA-TLX total workload} (mean = 24.53, SD = 11.99) compared to No-AI Summary condition. (mean = 27.40, SD = 15.72) (Figure \ref{fig:nasaTLX}). Despite the $\approx$ 3.9-point reduction, the Wilcoxon signed-rank test \cite{wilcoxon_individual_1992} did not indicate significant differences, with the same results for all subscales.

\captionsetup{justification=centering}

Although workload differences were not statistically significant, which may also be related to the small-N exploratory nature of this study, qualitative findings consistently highlighted perceived efficiency and sensemaking value, \hltext{suggesting benefits that may emerge more clearly in larger-scale deployments}. Importantly, introducing AI features did not significantly increase workload or reduce usability, suggesting that LLM augmentation can be integrated without adding a strongly notable interactional burden, while still being perceived as helpful. This robustness is noteworthy for HCI research on integrating AI into high-stakes workflows. 

Most participants reported that the summaries were valuable in reducing information overload and augmenting data sensemaking (N=15). Particularly, the summaries helped them reduce \textbf{time and effort} (N=6) \textbf{to understand the data, helped to give an overview of the patients, and easily compare with } WHO recommendations for physical activity \cite{who_cardiovascular_nodate}. In longer time spans (e.g., one year), summaries were seen as especially useful to quickly confirm trends. As one participant put it: \textit{“One of the things we are actually handling here is an overload of information … summaries do help simplify it”} (P12). Another added, \textit{“We don’t have a lot of time with the patient … it’s good to have this kind of helper”} (P2). Others echoed that summaries were most helpful under time pressure (P13) and that wearable data was otherwise “harder to analyze” (P6).

\subsubsection{I don't have to make a summary in my head: Summaries as Anchors for Data Exploration}

Summaries also changed how participants approached data exploration. Rather than scanning all charts first, many began with the summary and then checked specific graphs to confirm details. This offloaded the mental effort of “making a summary in my head” (P5). Another described, “\textit{I already know the summary so my eyes would be more focused on confirming it … I will spend less time and have more time for the patient}” (P2). 
Another participant highlighted that their use depended on the consultation goal (P4), “\textit{I think it would depend for me in terms of what the goal of the appointment was. So if this guy came from sedentary behavior only, maybe I would read this a little bit more than I would read other stuff.}” This shift resonates not only with distributed cognition but also with information foraging theory \cite{pirolli_sensemaking_2005}, as LLM-generated summaries altered "foraging cost structures" for participants by reducing the effort of scanning heterogeneous PGHD and directing attention toward high-value information.

\subsubsection{I don’t know if there’s any relationship: Desire for Correlations and Personalized Summaries}

Participants also expressed a desire for summaries to go beyond generic overviews and provide personalized narratives. In particular, they wanted summaries to highlight \textbf{correlations between data modalities} (e.g., links between physical activity and blood pressure) and to adapt to individual patient goals. As P15 suggested: \textit{“Maybe it would be interesting if there are some relationships … like the week he did less exercise, the blood pressure was worse.”} P6 emphasized the importance of tailoring summaries to specific clinical priorities, such as focusing on inactivity days rather than active ones,
\textit{“Maybe [it] could be good to be able to modify the summary to different patient… if you have the possibility of putting like one or two KPIs that you want, like some things you want to analyze in specific patients, and I think that [it] would be good to have a variable option on the summary}.”

Overall, participants envisioned summaries as tools that \textbf{align more closely with clinical reasoning}, offering adaptable outputs that reflect both patient context and professional workflow. This desire for personalization also shaped their experiences with the conversational interface, described in the next section.

\subsection{Value of Conversational Interfaces for Augmenting Workflow}

\subsubsection{We only have 20 minutes: Efficiency in Time-Limited Consultations.
}

Clinicians valued the conversational interface for its potential to save time in consultations, especially when working under 20-minute constraints. They described it as enabling faster access to patient metrics and supporting more personalized interactions. As one participant noted: \textit{“It’s very good … because in the consultation with the patients we have 20 minutes, and these two-[summaries and chat]-help me and the health professionals to do quality care. It’s important and personalized.”} (P7). Another added that while they were unsure about real-world uptake, the tool was clearly useful in time-limited settings (P16). P8 mentioned that with conversational interface, they would consider to skip manually reviewing the raw data altogether (P8). 

These reflections highlight that \textbf{efficiency, not just accuracy, influences adoption}, with conversational tools seen as a way of fitting longitudinal PGHD sensemaking into constrained clinical workflows.

\subsubsection{Augmenting Workflows; Personalizing Care}
Participants also described how conversational interfaces could \textbf{augment workflows and support more personalized care}. P5 explained how chat could help incorporate ad-hoc details that might otherwise be missed:  

\textit{“I would put a chatbot … where I would ask the information to the patient and I would put manually like some goals or some things that I observe … Maybe this would help because I'm having a trouble in my head systematizing every possible scenario for every possible personalization. [For example this] patient has some osteoarthrosis on his knee and he didn't give that feedback earlier so for me to add on what already is a clinical report.”} (P5)

P6 highlighted opportunities for comparing patients with similar profiles for better personalization:  

\textit{“I think a very good thing would be to tell me [about] three patients that are similar to this one. Like in profile but have different physical activity levels. And I think that can be useful. Like if I compare maybe, I don't know, two people that have almost the same life, but they manage to do very different levels of activity.”}

P4 described how conversational interfaces could generate both real-time graphs and tailored recommendations:  

\textit{“I think it will allow [me] to personalize the information we give to the patient even more. So according to what I know the patient needs and will work the best. I think it’s really cool to give me fast results … also give some tips in recommendation regarding other types of … physical activity or sedentary behavior and add that like to a report for example. ...“}

P4 further envisioned how the chatbot could allow for personalized lifestyle recommendations for patients based on their occupation,

\textit{And I think it would be really cool, for example, he says he's an engineer and … office worker. So it would be really cool to reduce your time doing work. And it gives like bullet points like try and do standing work, try and do your call while moving,… , try to go to the furthest bathroom when you need to break… [such recommendations] would be cool and besides the exercise plan to always give the patient like tips of how to decrease [sedentary] time through the day. And then … it can come here like on this side [gestures screen]“}

Others echoed that chat complemented dashboards (P3) and offered quick ways to understand patients during consultations (P13). Together, these reflections show how participants saw conversational interfaces as scaffolds for \textbf{workflow augmentation, personalization, and real-time patient engagement}, extending beyond what static dashboards could provide.

\subsubsection{We don’t have that knowledge: Bridging Data Literacy Gaps}
The conversational interface was particularly valued for \textbf{bridging gaps in data literacy}, enabling healthcare professionals without data-science expertise to conduct analyses they could not do themselves. P5 described it as filling a longstanding need:  

\textit{“The chatbot in the end I think that's the one of the best part of this tool! … In previous experience that we have in our team we were trying for a tool exactly like this and we had not had the resources for doing so … having the possibility to just ask for what I want to a chatbot … that really helps!”} (P5)

P4 emphasized how quickly the tool could generate visualizations and adapt goals:  

\textit{“I was impressed with how fast he did the things I was asking! … In less than one minute, I had this graph available … we don’t have the skills in terms of data analysis and using this type of code. We don’t have that knowledge.”} (P4)

These reflections highlight how participants saw conversational interfaces not only as efficiency tools but also as \textbf{equalizers}, democratizing access to analysis and visualization capabilities across HCPs with different data skills.

\subsection{Risks and Tensions in Adoption}
While participants recognized the value of AI in their workflow, many voiced concerns about professional boundaries, data accuracy and privacy, and (over) reliance on automation. In this section, we explore the tensions between efficiency and control, assistance and autonomy.  

\subsubsection{Assistant, Not Replacement or Not?: Ambivalence in AI autonomy preferences}
Participants envisioned AI supporting - but not replacing - their professional roles. Several highlighted how conversational tools could personalize recommendations or exercise prescriptions, yet stressed that final responsibility should remain with the healthcare professional. For instance, P15 envisioned capabilities beyond the prototype:  

\textit{“I have a question, for example, if [blinded for review] decides to play football once a week with his friends, can I ask, can you redo the program or not? … The best part I would like to interact is to have changes in the program … For example, I want him to have two days with high intensity activity.”} (P15)

Others emphasized that human expertise and social presence were irreplaceable. As P8 explained:  

\textit{“From the exercise physiologist, I could also just ask, OK, based on that person's profile, please fill me an exercise plan. And I could have it. But then there are some things that AI cannot replace—the social part, the potential of involving people … the physician.”}  - (P8)

Several participants (e.g., P9, P3, P16) drew firm boundaries around professional tasks, noting that AI could suggest physical activity tips but should not prescribe exercise, which requires screening and domain expertise. At the same time, two participants expressed how non-domain experts such as general practitioners with less training might benefit from AI support (P4, P8).  

\textit{“I think if it's someone like a doctor that has more difficulty interpreting this type of graphic and this type of data, this is very useful. So I would say for me, it wasn't that different because I can understand what it's saying. But I would say for sure if it's another professional working in this field or giving recommendations or general practitioners, this would be very, very, very good. “} - \hltext{(P4)}

Despite this ambivalence, there was broad agreement that AI should act as an \textbf{assistant rather than a replacement}. As P2 summarized:  

\textit{“I think AI is not going to replace our jobs but they are going to help us to reduce the time that we spend … they will be like our helper.”} (P2)

It is important to note that the prototype used in the study did not generate exercise prescriptions or lifestyle recommendations. When participants described such possibilities, they were articulating their expectations and aspirations for future systems rather than experiences with our tool. 

\subsubsection{Overreliance and Risk of Deskilling}
Participant confidence in their final plans - measured on a 1-5 self-assessed Likert scale where 5 refers to most confident - was comparable across conditions (AI: \textit{M}=4.38, \textit{SD}=0.67; No-AI: \textit{M}=4.27, \textit{SD}=0.68). Trust in AI summaries - measured using a self-assessed Likert scale from 1-5 where 5 indicates full trust - was relatively high (M=4.27, SD=0.98), and correlated with confidence (Spearman r=.46, p=.001), suggesting reliance increased with trust. Higher trust in the AI summary was therefore linked to greater confidence in the final physical activity plan. This suggests that HCPs who trusted the summaries were also more likely to rely on them in shaping their plans. Our qualitative insights contextualize this relationship: while participants valued the support of summaries, they emphasized the risks of “blind trust”.

As P12 noted: \textit{Once we have them, we will trust them, but we need to trust them in the right way so we're not...putting all of our confidence into these specific summaries and then .. I will be misinterpreting these results.”}. Similarly, P14 highlighted the danger of overreliance: \textit{“… with recommendation, the risk can be trusting a hundred percent and not thinking about it, just [asking the chatbot] what should I recommend to this patient.”}  

P8 went further, raising concerns about professional deskilling: \textit{“People are lazy in general … I’m afraid that people are just using it for everything … everything could be answered in just one or two words or just asking the phone and quite like an Alexa or whatever… , I think that most of the people that are not aware, they will just use it without stimulating creativity because you can still be a lot of creative, even more creative with this… I would say, for sports scientists, [AI] can probably turn them a little bit lazier.”} 

Finally P6 also cautioned that real-time, automatically generated outputs could create anxiety for patients themselves, particularly if they were not prepared to interpret such feedback. This extends concerns about overreliance from professionals to patients, underscoring the need for careful mediation of AI outputs in clinical encounters.

Together, these reflections highlight concerns that while AI can boost confidence and efficiency, it also risks \textbf{automation bias and erosion of professional judgment}, requiring careful safeguards in clinical practice.

\subsubsection{What’s behind it?: Accuracy, Transparency, and Trust Concerns}
\label{sec:accuracy-transparency-trust}
Participants linked trust to both the accuracy of outputs and transparency about how summaries were produced. Several noted that while they valued the tool, AI is not always reliable and requires critical oversight. As P16 explained:  

\textit{“I mean, I think the AI right now is really prepared to analyze this kind of simple data. It's just about means and means graphs … For descriptive statistics, I trust a lot on AI … but for correlations, I would probably not ask AI.”} 

P12 spotted a mistake in one summary but emphasized its usefulness overall:  

\textit{“I think the summary actually helps and in God we trust, although when I actually saw the whole blood pressure [summary error], I think [the LLM] needs to be a little bit more fine-tuned so we don’t have things like, oh it’s stage two or stage one … when a person is not actually even within the cut-off points.”} (P12)

P4 suggested safeguards such as disclaimers to flag possible errors, \textit{“I think the risk is if it gives me wrong information… it didn't tell me like beware, data might be wrong or something, maybe it would induce me in error if I didn't have the perception that that was a valid… So should there be some indication to say this is AI generated?”}

Meanwhile P6 noted their own limited understanding of LLM capabilities, \textit{“AI [has] the same risks as ChatGPT. We need to be critical about it … Actually, I don’t know [about] ChatGPT enough to know all the errors it can have.”} 

For some participants, trust was also described as context-dependent. P11 felt generally comfortable with trusting the AI outputs despite potential mistakes:  
\textit{“The algorithm of the AI, I think, can induce some mistakes, I guess. But if I feel that it is trustable, I wouldn’t be afraid of it.”} (P11)

P4 distinguished between using chat to interpret data versus asking for generic recommendations, \textit{“Because if I’m asking the chat to tell me something about this data, maybe I will trust it a little bit more. If I’m asking … about the recommendations, I have to be aware that it can be with mistakes.”}

P13 also suggested clickable links from summaries to charts (P13) to improve transparency of how the summaries were generated and trustworthiness of the data, "{Click [the summary] and it goes to the information [Gestures on chart] . Okay, that's a center average six point seven hours… [this could improve trust towards the summaries]}“

These reflections highlight how participants balanced enthusiasm with caution, linking trust to \textbf{accuracy, transparency, provenance, and their own AI literacy}. 

\subsubsection{It’s scary, it's [like] talking to someone I don't know: Privacy Risks}
Only a few participants raised explicit concerns about privacy. P9 compared entering personal details into the system to \textit{“talking to someone I don’t know”}, describing unease about where the data might go. P10 similarly emphasized that health data must be strongly protected:  

\textit{“The bad thing is that this can be a target … for data protection. You know, someone can come here and take this data. So this should be secure enough so people can trust to put their data here.”} (P10)

Beyond security, participants also highlighted the need for \textbf{transparency about data flows}—where information goes once it is entered and how it is processed by AI. For them, uncertainty about these “black box” pathways was itself a privacy risk, reinforcing the need for clear communication about data security and governance while implementing AI systems in healthcare.

These comments, though limited in number, underscore the importance of \textbf{transparent data flows and strong data and AI governance strategies} to build trust in clinical adoption.

\subsubsection{Age related technology acceptance barriers}
Although our sample was relatively young (23–42 years), several participants perceived that \textbf{older colleagues might be less inclined to adopt AI tools}. P1 remarked: \textit{“The older colleagues, I think it’s more difficult because [of] the barrier from the new technology, but young colleagues … yes totally sure.”} Others echoed that younger HCPs would likely adapt more easily, while older colleagues might struggle with understanding how such AI-based systems worked(P3, P6). These reflections represent perceptions rather than evidence, and also highlight the risk of reinforcing stereotypes. They underscore the importance of \textbf{age-inclusive design and avoiding digital ageism} in clinical AI adoption.

\section{Discussion}
\hltext{Recent HCI and clinical informatics work has shown how LLMs can generate clinically relevant narratives from personal and behavioral health data-e.g., in mental health journaling} \cite{kim2024}, \hltext{ mobile sensing interpretation }\cite{Englhardt24}, \hltext{and personal health informatics workflows }\cite{shaan2025} \hltext{and even outperform HCPs in summarizing information from electronic health records} \cite{van2024adapted}. \hltext{Complementing these patient- and data-centric perspectives, recent HCP-facing research on LLM integration into clinical workflows highlights that HCPs generally welcome AI augmentation while simultaneously emphasizing the importance of appropriate AI-governance} \cite{fraile2025understanding}.

Guided by our research questions, we explored: \textbf{1)} how LLM summaries influenced HCPs' sensemaking of health data, \textbf{2)} their perceived benefits and acceptance of LLMs for integrating them into their workflows, \textbf{3}) their needs and interests regarding using a conversational interface to interact with PGHD to augment \hltext{cardiac risk reduction} workflows. 

Our findings illustrate how LLMs can act as \textit{cognitive sensemaking partners} in augmenting rather than replacing professional expertise, echoing Hutchins' notion of cognition distributed across people and artifacts \cite{hutchins_cognition_1995} and extending HCI debates on how sociotechnical systems scaffold human judgment under time pressure \cite{shneiderman_human-centered_2020}. At the same time, findings highlight professionals' concerns around accuracy, ethical boundaries, and adoption barriers. These findings inform not only clinical PGHD workflows but also broader HCI contexts where professionals face data overload and heterogeneous literacy, contributing to theory and design of AI-augmented data sensemaking.

To translate these findings into actionable design implications, we suggest a set of design implications (DI) (see Tables \ref{tab:ai-augmentation}, \ref{tab:ai-autonomy} and \ref{tab:trust-overreliance}). We reference these implications throughout the discussion using DI numbers with a prefix based on sections (e.g., DI-A1 - DI-A3), while the tables provide expanded descriptions. \hltext{Our findings provide formative evidence on how AI-generated summaries and conversational interfaces may function as cognitive sensemaking partners for HCPs reviewing multimodal PGHD. Rather than assessing clinical effectiveness or system accuracy, our goal was to explore perceptions, workflow fit, and sociotechnical considerations that would shape responsible integration of AI augmentation in practice. The results highlight both the potential of AI to scaffold data sensemaking and the tensions—particularly around trust, agency, transparency, and overreliance—that must be addressed for real-world adoption.}

\subsection{AI’s Augmentation Potential: Summaries and Chat as Cognitive Partners}
Our findings suggest that HCPs generally perceived LLMs as \hltext{\textit{collaborative data science assistants rather than their substitutes to their profession.}} In the context of cardiac risk reduction, multi-modal wearable data (such as blood pressure, step counts, physical activity minutes, sedentary time, sleep duration etc.) which could give a holistic and objective picture of patients, but result in \textbf{information overload}. As recommended by the American Heart Association\cite{golbus_digital_2023}, AI summaries or insights could automate interpretation of data and reduce load on HCPs, enabling them to provide personalized care for patients \textbf{(DI-A1)}.

Prior work has shown that visualization literacy and clinical decision-making ability are significantly related \cite{weil_visualization_2022} and that HCPs’ insufficient expertise in working with data is a barrier to integrating PGHD in clinical workflows \cite{west_common_2018}. As highlighted by P4, HCPs might not necessarily have the data science skills or knowledge to perform quick visualizations and analysis. Our prototype shows the potential of leveraging LLMs to synthesize PGHD into insights and bridge HCPs’ visualisation literacy gaps \textbf{(DI-A2, DI-A6)}.

Reflecting on the AI features in the dashboard from an HCI lens, the AI-enabled summaries and conversational interface can be framed as HCPs’ \textbf{cognitive sensemaking partner} who helps offload the mental effort of “making a summary in [their] head” (P5). Within the \textbf{distributed cognition} framework \cite{hollan_distributed_2000}, the LLM summaries position themselves as an \textbf{ external cognitive resource} that aims to help offload HCPs’ tasks related to data sensemaking \cite{grinschgl_supporting_2022} thereby improving understanding of the patient, resulting in better decision making.

This positions LLM-augmented sensemaking not only as workflow support but as an enabler of behavior change dialogues, scaffolding shared decision-making and motivational framing, extending existing work on behavior change support and adaptive health technologies \cite{fang_physiollm_2024, wurhofer_investigating_2024, gartner2025digitally}.

The ability to have personalized data analysis and charts to show patients - e.g. “\textit{Look at how much sedentary time you have compared to physical activity}” or “\textit{You can see that your sleep and physical activity are correlated}” could help with scaffolding dialogues to enhance shared decision making \textbf{(DI-A3, DI-A4)}. Moreover, as mentioned by one of the HCPs, real-time analysis and chart generation capabilities comparing a patient's data with population data could help to educate patients in a personalized way about physical activity behaviour change. As described by Pedooras and Khera in their review \cite{pedroso_leveraging_2025} on how AI could enhance cardiac care, such AI-assisted data interpretation could allow HCPs to spend less time performing data interpretation [and analysis] and more time with patients.

Participants further imagined how LLMs could augment creation of personalized recommendations, modification of exercise plans and generation of summaries for patients - tasks considered as manual and time-intensive \textbf{(DI-A5)}. This aligns with several developments in the healthcare industry, which focuses on using AI tools such as AI scribes \cite{van_buchem_digital_2021} to reduce administrative burden among HCPs. Over time, there is a potential for system designers to analyse HCPs' interaction with the system to improve the queries and dashboard design to adapt with their reasoning patterns \textbf{(DI-A7)}.

Overall, with data science literacy gaps, time pressure and overwhelming yet useful data, HCPs envisioned LLMs as a collaborative data science assistant that augments cognitive capacity, supports personalized patient engagement, and reduces burden - while having HCP oversight and expertise. 

LLMs summaries have the potential to augment HCPs' data sensemaking capabilities by acting as cognitive scaffolds \textbf{(DI-A1)} - where scaffolding refers to a process that enables a novice to solve a problem, carry out a task or achieve a goal which would be beyond his unassisted efforts \cite{wood1976role}. Furthermore, the conversational interaction with LLMs for data exploration and analysis acts as a potential visualization literacy equalizer, allowing HCPs to have better capabilities in understanding their patients \textbf{(DI-A6)}. Over time, systems can adapt to recurring query patterns and reasoning strategies also raising opportunities for dashboards that evolve with clinical practice \textbf{(DI-A7\hltext{)}} rather than remaining as static tools.

\begin{table*}[t]
\centering
\caption{Design Implications for AI Augmentation (A)}
\label{tab:ai-augmentation}
\begin{tabularx}{\textwidth}{p{0.08\textwidth}X}
\toprule
\textbf{\#} & \textbf{Design Implication} \\
\midrule
DI-A1 & \textbf{Use summaries as cognitive scaffolds}: AI-generated summaries should act as glanceable cognitive scaffolds \cite{jones_sensemaking_2016,scholich_augmenting_2024}, reducing information overload while enabling HCPs to rapidly grasp longitudinal PGHD. \\
DI-A2 & \textbf{Support dual interaction modes}: Combine low-effort glanceable summaries with higher-effort natural language explorations. This scaffolds flexible sensemaking and supports diverse workflows and preferences. \\
DI-A3 & \textbf{Allow dynamic and personalized adjustment of summary focus}: Interfaces should enable clinicians to personalize the LLM's analytical focus, e.g., prioritizing inactivity days or flagging risk-relevant patterns, thereby aligning AI output with professional judgment. This could be implemented via a personal library of customized prompts.\\
DI-A4 & \textbf{Enable cross-modal data exploration to align with HCP analytical workflows}: Support questions like “Compare sedentary time with BP last week,” allowing HCPs to surface clinically meaningful patterns across modalities. \\
DI-A5 & \textbf{Support summary personalization}: Summaries should be personalizable to reflect patient-specific goals and conditions, allowing HCPs to define which metrics or patterns are emphasized based on clinical context. \\
DI-A6 & \textbf{Embed natural language chat for customizable data exploration}: Conversational interfaces should allow HCPs to ask context-driven questions and receive tailored summaries or visual outputs without needing data analysis expertise. \\
DI-A7 & \textbf{Improve LLM model and dashboard over time}: Learn from aggregated HCP queries, enabling the system to suggest relevant visualizations or surface commonly requested data views. For example, if many HCPs consistently ask about correlations between sedentary time and physical activity, this insight could inform dashboard redesigns and suggested insights to align the interface more closely with HCPs’ reasoning patterns.\\
\bottomrule
\end{tabularx}
\end{table*}

\subsection{Boundaries of Autonomy: Role-Based Acceptance}

The participants primarily saw AI as a collaborative partner in achieving their tasks rather than a substitute in line with comments by Sezgin \cite{sezgin_artificial_2023}. Two participants (P8 and P3) explicitly said that sports scientists should use the tool for tasks apart from solely prescribing exercise - a key skillset they possess. These time-intensive tasks include generating personalized lifestyle change suggestions (e.g. based on occupation, demographics, generating summary reports for patients to keep). Such tasks are suitable for \textbf{intelligent automation} \cite{tan_rip_nodate}. 

On the other hand, some HCPs acknowledged that non-domain experts, such as general practitioners and cardiologists whose primary roles are not in exercise prescription or have reduced time for crafting exercise prescription, could benefit from using such AI-augmentation with discretion. 

We found that the acceptance of AI shifts depending on the healthcare professional’s role and context. This is consistent with existing research demonstrating that different HCP specialties have unique needs, necessitating specific decision-support system designs \cite{serban_i_2023}. Furthermore, it reflects the ongoing tension between AI as an augmenting or a substituting technology, which directly impacts clinicians’ autonomy and authority \cite{kim_how_2024}. 

However, when integrating AI into the healthcare triad, it's necessary to consider patient perspectives to avoid a 'third wheel' effect \cite{triberti_third_2020}, where the AI's presence could influence the attitudes of both doctors and patients during decision-making. Kim \textit{et al.} \cite{kim_how_2024} found that patient preferences for AI autonomy vary based on perceived risk in decision making, individual characteristics such as attitudes towards healthcare decision making and even over time. 

\hltext{Although prior work has assessed clinicians’ preferred degrees of automation and advocated for graduated AI autonomy} \cite{fraile2025understanding}, \hltext{our findings on HCP preferences for role based AI-autonomy show a need for future work investigating HCP perspectives on the ethical limits and patient risk/safety considerations concerning the role of} AI \textbf{(DI-B1)} \hltext{along different specialties in a cardiac risk reduction context, and possibly extending towards patient perspectives to reduce tensions between both parties in triadic setups involving AI, HCP and patients.}

Ultimately, when designing AI integration into the system, it is important to consider the system autonomy levels \cite{fleming_lindsley_haba-maba_2021}. However, in the healthcare domain, AI integration becomes complex due to medical ethics, patient safety and the dynamics between HCPs and patients. \hltext{The HABA-MABA framework - “humans are better at – machines are better at”}, or \textit{HABA-MABA} \hltext{from human factors research identifies complementary human and machine strengths. AI excels at enabling the rapid processing of longitudinal, multimodal data }\cite{zhang_sensorlm_2025},\hltext{ while humans contribute irreplaceable social and empathic capabilities }\cite{kim_how_2024},\hltext{ a division of labour that participants implicitly endorsed when positioning AI as assistant rather than replacement}. System designers have to consider the “goals of the system” to clearly elucidate the 1) levels of autonomy of the system, 2) human-driven tasks in case of the lowest level of automation i.e. no automation \cite{fleming_lindsley_haba-maba_2021} with patient safety and risk minimization as a key goal \textbf{(DI-B1)}. Unlike automation and aviation, which have clearly defined automation levels ranging from no automation to full automation, there are currently no standardized levels of AI autonomy in medicine \cite{bitterman_approaching_2020}. An in-depth understanding of the clinical context and real-world pilot testing with HCPs in controlled settings paves the path towards implementation \textbf{(DI-B2)}. Such efforts in turn allow HCPs to establish trust in the system and also ease adoption and acceptance \cite{lambert_integrative_2023}.

\begin{table*}[t]
\centering
\caption{Design Implications for AI Autonomy (B)}
\label{tab:ai-autonomy}
\begin{tabularx}{\textwidth}{p{0.08\textwidth}X}
\toprule
\textbf{\#} & \textbf{Design Implication} \\
\midrule
DI-B1 & \textbf{Investigate role-sensitive AI autonomy}: Medico-legal standards should be established to determine configurable levels of AI autonomy for generalists and specialists in the context of cardiac risk reduction. This could allow experts to have AI augmentation rather than replacement, preserving their control in exercise prescription and decision support. \\
DI-B2 & \textbf{Include end-users in AI development and integration lifecycle}: Involving end users in the development and integration process could allow for feedback loops to better integrate AI into their clinical workflows, thereby improving trust and acceptance. \\
\bottomrule
\end{tabularx}
\end{table*}

\subsection{Risks and Implementation Factors}

While participants recognized the value of AI summaries and conversational interfaces, they also raised concerns about potential risks, particularly \textbf{overreliance}. We found that when participants trusted AI summaries more, they felt more confident in the final activity plan they created. These align with recent experimental work showing how higher trust can amplify reliance and that HCPs might lean more heavily on AI in a high-stakes context to offload responsibility and cognitive effort \cite{fahnenstich_trusting_2024}. Putting together the qualitative and quantitative findings, we observe tensions where under high stakes or high-pressure situations, there are increased changes of \textbf{overreliance} and potential risks for \textbf{patient safety}. With overreliance, there are also potential long-term risks to human cognition, such as a decline in the HCPs' own analytical skills due to over-reliance on the tool \cite{grinschgl_supporting_2022} or laziness as quoted by P8. This could particularly be of higher risk, in Low- and Middle-Income Countries (LMIC) countries or locations, where patient/HCP ratio might be high or where the healthcare system is overburdened. The potential of automation bias and mitigating them via safeguards and warnings are key for safety \textbf{(DI-C3 and DI-C4)}. 

HCPs also felt that they needed to know the source of the data used for the algorithms as well as the transparency of the algorithm to improve their trust and acceptance of the AI summaries \textbf{(DI-C1)}. In addition to that, they acknowledged that AI has its limits, resulting in uncertainty over how it works behind the hood - “black box” and are unaware of how the data flows \textbf{(DI-C2)}. In a related study on exploring the use of AI summary to help with data sensemaking of glucose readings, the authors found that participants decided not to look at the algorithmic outputs due to concerns about contextual integrity, accuracy of the algorithms \cite{scholich_augmenting_2024}. Echoing earlier findings on the value of visualization support, one participant suggested placing summaries directly alongside data points to improve trust, aligning with prior work \cite{scholich_augmenting_2024}.
Improving transparency relates to the theme of legibility - making data and analytics algorithms both transparent and comprehensible to the people the data and processing concerns - as framed in the concept of human-data interaction proposed by Motier \textit{et al.} \cite{mortier_human-data_2014}. 
Yet, such explanations, which improve legibility or transparency, can potentially increase “cognitive labour and compromise critical engagement” with the data \cite{scholich_augmenting_2024}.

There is also a need to educate HCPs on AI to allow understanding of limitations - that AI can make mistakes - and to let them know that they should always give the final judgement and are responsible for prescriptions (DI-C6). As quoted by one of the participants, \textbf{warning messages such as “}\textit{\textbf{This output was generated by AI. AI can make mistakes. You should make the final decision.}} ” which could in turn reduce potentials of overreliance amongst HCPs \textbf{ (DI-C7)}.

Finally, participants' age is an adoption barrier. These echo prior work on inter-generational barriers to adoption in healthcare technology\cite{nakagawa_inter-generational_2020}. This implies that more interventions and training/reskilling could be provided to increase adoption of AI tools in their workflows responsibly \textbf{(DI-C11)} . Additionally, systems could be made personalizable or reconfigurable based on individual abilities and preferences \cite{branco_co-designing_2024} \textbf{(DI-C5)}.

\hltext{Together, these reflections surface a broader tension: while AI can boost confidence and efficiency, it simultaneously risks fostering automation bias and eroding professional judgment -- what Navarro \textit{et al.} term the "automation paradox" in clinical AI adoption} \cite{fraile2025understanding}. These concerns point to concrete design challenges for HCI: e.g., \hltext{provenance-linked  summaries} \textbf{(DI-C1)} \cite{EU_AI_Act_2024}, transparency mechanisms tailored to clinical contexts \textbf{(DI-C1 and DI-C2)}.

\begin{table*}[t]
\centering
\caption{Design Implications to Reduce Risks and Ease Implementation (C)}
\label{tab:trust-overreliance}
\begin{tabularx}{\textwidth}{p{0.08\textwidth}X}
\toprule
\textbf{\#} & \textbf{Design Implication} \\
\midrule
DI-C1 & \textbf{Improve trust towards summaries through data provenance}: Summaries should be traceable, e.g., clicking on a summary sentence should highlight the corresponding data to build trust and help HCPs verify LLM outputs quickly. \\
DI-C2 & \textbf{Clarify data processing}: Systems should inform users where data is processed and stored (e.g., “processed locally on hospital server”) to alleviate privacy concerns. \\
DI-C3 & \textbf{Use warnings to ensure oversight}: Warnings such as “This output was generated by AI. AI can make mistakes. You should make the final decision,” should be used to reinforce clinician oversight and reduce risks of overreliance or automation bias. \\
DI-C4 & \textbf{Provide HCPs usage-based overreliance indicators}: Visual cues such as engagement gauges or summaries of how often the AI summary is used can help clinicians track their dependence and prompt caution. \\
DI-C5 & \textbf{Support reconfigurable interface complexity}:  Offer flexible modes such as \emph{Novice} vs. \emph{Expert} to align with user skills and preferences.  \\
DI-C6 & \textbf{Provide continual AI education and literacy support}: Structured educational materials and training sessions should help HCPs understand AI limitations, appropriate use, how to interpret outputs safely and integrate them into their workflows to improve acceptance\\

\bottomrule
\end{tabularx}
\end{table*}
\subsection{Limitation}
\label{limit}
The LLM was not pre-trained on sensor data, so the summaries are not the most accurate . Future work can integrate fine-tuned LLMs into the pipeline. Because the study sessions did not impose strict time limits, any potential efficiency gains from the AI summaries may have been muted in the quantitative workload and confidence measures. The participant pool (16 clinicians) was adequate for initial insights but underpowered for detecting small or medium effects in non-parametric tests; larger samples may reveal subtler quantitative differences. Our study relied on synthetic PGHD generated for stratified personas rather than real-world patient data. This approach allowed us to control complexity across participants and avoid privacy risks, but it was also necessitated by the lack of available, longitudinal multimodal PGHD datasets suitable for research in this context. \hltext{Synthetic data cannot fully capture the variability, noise, or missingness typical of real-world PGHD, and findings should therefore be interpreted as reflective of controlled conditions rather than deployment realities. We evaluated the prototype only with HCPs working alone, rather than in real-time with patients present. While this allowed us to focus on clinicians’ sensemaking processes without introducing additional ethical and logistical constraints, it also means that we do not yet capture how AI summaries and conversational interfaces would shape triadic interactions during live consultations.} The findings should be interpreted as insights into HCPs’ perceptions and interactions under controlled conditions. Future work will extend this evaluation to real, longitudinal PGHD in collaboration with clinical partners once appropriate datasets become accessible. Also, AI has been known to make developers even slower \cite{becker_measuring_2025} and \hltext{studies have shown heterogeneity in the effects of human-AI collaboration }\cite{vaccaro_when_2024}. As such, there needs to be studies to investigate Human-AI synergy while making sense of data. More research is needed on how to present AI summaries effectively, as oversimplified natural language / conversational interface presentation can lead to misinterpretation risks due \cite{henzler_healthcare_2025}. Future work should also \hltext{investigate} HCP perspectives on role-based AI autonomy and levels of autonomy for physical activity recommendations and prescriptions.

\subsection{Conclusion}
\label{conclusion}
We embarked on a mixed-methods study to investigate LLM integration to support HCPs with data sensemaking and data exploration of PGHD in the context of cardiac risk reduction.  Our study demonstrates how LLM summaries and conversational interfaces can reconfigure data sensemaking for clinicians, acting as \textit{cognitive sensemaking partners} in time-pressured health contexts. These insights extend beyond cardiac care to broader HCI debates on AI-augmented data interaction, and the design of trustworthy, literacy-bridging analytic systems.
However, HCPs cautioned about risks of AI integration, such as algorithmic errors and the need for transparency over data privacy, security, algorithms, potential for overreliance and deskilling, and varied preferences for AI autonomy for exercise recommendations and prescriptions. We share concrete design implications on addressing barriers and enablers for AI integration to support HCPs with data sensemaking, highlight the critical need for future work to investigate AI autonomy and controlled experimental studies to evaluate if AI indeed augments data sensemaking and decision making.

\clearpage
\bibliographystyle{ACM-Reference-Format}
\bibliography{bibliography}

\newpage
\newpage
\appendix
\section{Appendices}
\label{appendix}
\subsection{Interview Guide}
\label{app:interview}
You’ve just seen a prototype of a large language model (LLM)-powered tool that can generate natural language / conversational interface summaries based on patient-generated health data and also the ability to interact with the data using natural language / conversational interface. Our aim is to understand how such a tool might eventually support your work with patient data in physical activity planning during cardiac rehabilitation.

\subsection*{Experiences with the LLM Tool (RQ1: Sensemaking)}

\textbf{Main Question:}
\begin{itemize}
    \item What was your impression of the summaries generated by the tool?
    \item How did you find working with them besides looking at the data?
    \item How did the LLM-generated summary influence your interpretation of the data?
    \item Did the LLM summaries highlight anything you might have missed if you'd only looked at the data?
    \item How did you feel about being able to interact with the chatbot to interact with the data?
\end{itemize}

\textbf{Follow-ups:}
\begin{itemize}
    \item Was anything surprising or unexpected?
    \item Compared to how you typically work with patient-generated data, how did this approach differ in terms of usability, decision-making?
\end{itemize}

\subsection*{Reflections on Value and Fit (RQ2: Costs and Benefits)}

\textbf{Main Question:}
\begin{itemize}
    \item From your perspective, what stands out as something that could be valuable about this tool?
    \item What could be problematic about using a tool like this?
    \item What do you think about patients having access to such a tool?
\end{itemize}

\textbf{Follow-ups:}
\begin{itemize}
    \item What aspects of the tool seem promising or concerning?
\end{itemize}

\subsection*{Broader Considerations and Expectations (RQ3: Acceptability)}

\textbf{Main Question:}
\begin{itemize}
    \item Thinking about your own clinical practice, how do you see a tool like this fitting in?
\end{itemize}

\textbf{Follow-ups:}
\begin{itemize}
    \item What would make it easier or harder for you to trust or use something like this?
    \item How do you think your colleagues might react to a tool like this?
    \item What changes or additions would make this tool more useful to you?
    \item What are your perspectives on getting recommendations from this tool?
\end{itemize}

\subsection*{Wrap-up}
\begin{itemize}
    \item Is there anything else you want to share about your experience using this system?
\end{itemize}

\newpage
\subsection{Prompt}
\label{app:prompt}
\appendix
\section{ChatGPT Configuration}

\subsection*{Model Settings}
\begin{MyVerbatim}
Model: gpt-4-turbo
Temperature: 0.5
Max tokens: 1024
\end{MyVerbatim}

\subsection*{Summary of Prompt and Data Passed}

\textbf{Physical activity}
\begin{MyVerbatim}
You are a health data analyst specializing in summarizing physical activity data 
for a specific time period. 

Provide objective analysis of activity data without making specific recommendations. 
Focus on patterns, trends, anomalies, maximum and minimum values relative to WHO guidelines.
Don't make assumptions. Only make observations from the provided data.

Format your response using Markdown for better readability (e.g., use **bold** for important values,
use bullet points for listing items, use headings for sections).

Use a neutral tone.  
Keep your response concise (3-4 sentences and within 200 characters).

Column Info:
Date: Date of the recorded data (YYYY-MM-DD)
LPA: Minutes of light physical activity per day
MPA: Minutes of moderate physical activity per day
VPA: Minutes of vigorous physical activity per day

Data passed:
Patient Name { }
Time Frame { }
Data { }
\end{MyVerbatim}

\textbf{Sedentary Time}
\begin{MyVerbatim}
You are a health data analyst specializing in summarizing sedentary time data 
for a specific time period. The sedentary time is input as minutes, however, 
the output analysis should be in hours with one decimal place.

Provide objective analysis of sedentary data without making specific recommendations.
Focus on patterns, trends, anomalies, maximum and minimum values.
Don't make assumptions. Only make observations from the provided data.

Format your response using Markdown for better readability.
Use a neutral tone.
Keep your response concise (3-4 sentences and within 200 characters).

Column Info:
Date: Date of the recorded data (YYYY-MM-DD)
Sedentary: Minutes spent in sedentary activity per day

Data passed:
Patient Name { }
Time Frame { }
Data { }
\end{MyVerbatim}

\textbf{Blood Pressure}
\begin{MyVerbatim}
You are a health data analyst specializing in summarizing blood pressure readings 
for a specific time period.

Provide objective analysis of blood pressure data without making specific recommendations.
Focus on patterns, trends, anomalies, maximum and minimum values.
Don't make assumptions. Only make observations from the provided data.

REFERENCE RANGES:
Normal: <120/<80 mmHg
Elevated: 120-129/<80 mmHg
Stage 1 Hypertension: 130-139/80-89 mmHg
Stage 2 Hypertension: >=140/>=90 mmHg

Format your response using Markdown for better readability.
Use a neutral tone.
Keep your response concise (3-4 sentences and within 200 characters).

Column Info:
Date: Date of the recorded data (YYYY-MM-DD)
Systolic_BP: Systolic blood pressure (mmHg)
Diastolic_BP: Diastolic blood pressure (mmHg)

Data passed:
Patient Name { }
Time Frame { }
Data { }
\end{MyVerbatim}

\textbf{Sleep}
\begin{MyVerbatim}[breaklines=true, breakanywhere=true]
You are a health data analyst specialized in summarizing sleep patterns 
for a specific time period.

Provide objective analysis of sleep data without making specific recommendations.
Focus on patterns, trends, anomalies, maximum and minimum values.
Don't make assumptions. Only make observations from the provided data.

Format your response using Markdown for better readability.
Use a neutral tone.
Keep your response concise (3-4 sentences and within 200 characters).

Column Info:
Date: Date of the recorded data (YYYY-MM-DD)
Total_Sleep_Duration: Total hours of sleep per night
Light_Sleep: Hours of light sleep per night
Deep_Sleep: Hours of deep sleep per night
REM_Sleep: Hours of REM sleep per night
Wake_Time: Number of times the participant woke up at night
Sleep_Quality: Sleep quality score (1-5, higher is better)

Data passed:
Patient Name { }
Time Frame { }
Data { }
\end{MyVerbatim}

\textbf{Combined Data}
\begin{MyVerbatim}
You are a health data analyst AI assisting a healthcare professional (doctor) 
who is summarizing a specific patient's combined health and physical activity data 
(activity, sedentary, blood pressure, sleep) for a specific time period. 

Focus on patterns, trends, anomalies, maximum and minimum values relative to WHO guidelines. 
The sedentary time is input as minutes, however, the output analysis should be in hours 
with one decimal place.

Your role is to help the doctor understand this patient's data and provide holistic, 
data-driven, clinically relevant insights based on the doctor's questions without 
making any recommendations.

Do not make specific medical recommendations or diagnoses.
Format your response using Markdown for better readability.
Use a neutral tone.
Keep your response concise (3-4 sentences and within 200 characters).

Column Info:
Date: Date of the recorded data (YYYY-MM-DD)
MPA: Minutes of moderate physical activity per day
VPA: Minutes of vigorous physical activity per day
Sedentary: Minutes spent in sedentary activity per day
Systolic_BP: Systolic blood pressure (mmHg)
Diastolic_BP: Diastolic blood pressure (mmHg)
Total_Sleep_Duration: Total hours of sleep per night
Light_Sleep: Hours of light sleep per night
Deep_Sleep: Hours of deep sleep per night
REM_Sleep: Hours of REM sleep per night
Sleep_Quality: Sleep quality score (1-5, higher is better)

Data passed:
Patient Name, age, gender and diagnosis { }
Time Frame { }
Data { }
\end{MyVerbatim}

\newpage
\textbf{\hltext{Provenance Analysis of LLM outputs}}

\begin{table}[h]
\centering
\small
\begin{tabular}{lccc}
\toprule
\textbf{Modality} & \textbf{\# Instances} & \textbf{MAPD (\%)} & \textbf{Median (\%)} \\
\midrule
MPA (minutes) & 31 & 10.72 & 1.98 \\
VPA (minutes) & 26 & 10.46 & 2.54 \\
Sedentary Time & 35 & 1.72 & 0.00 \\
BP (Systolic) & 24 & 0.76 & 0.24 \\
BP (Diastolic) & 24 & 0.41 & 0.21 \\
Sleep Duration & 31 & 0.52 & 0.31 \\
Sleep Quality & 13 & 1.58 & 3.33 \\
\midrule
\textbf{Totals} & 184 & \textbf{3.96} & \textbf{0.47} \\
\bottomrule
\end{tabular}
\caption{Breakdown of provenance accuracy across modalities - holistic insights.}
\label{tab:summaryProvenance}
\end{table}

\begin{table}[h]
\centering
\small
\begin{tabular}{lccc}
\toprule
\textbf{Modality} & \textbf{\# Instances} & \textbf{MAPD (\%)} & \textbf{Median (\%)} \\
\midrule
Sleep Duration & 4 & 0.57 & 0.19 \\
Sleep Quality & 1 & 0.00 & 0.00 \\
MPA (minutes) & 7 & 4.82 & 2.87 \\
VPA (minutes) & 9 & 1.45 & 1.57 \\
MVPA (minutes) & 2 & 4.94 & 4.94 \\
Sedentary Time & 7 & 3.04 & 3.13 \\
\midrule
\textbf{Combined} & 30 & \textbf{2.68} & \textbf{1.57} \\
\bottomrule
\end{tabular}
\caption{Breakdown of provenance accuracy across modalities - chat logs.}
\label{tab:chatProvenance}
\end{table}




\end{document}